\documentclass[twocolumn,aps,amsmath,amssymb,superscriptaddress,prb]{revtex4-1}   \usepackage[latin2]{inputenc}
\usepackage{amsmath}
\usepackage{amsfonts}
\usepackage{bm}
\usepackage{dcolumn}
\usepackage{setspace}
\usepackage{graphicx}
\usepackage{color}
\usepackage{multirow}
\usepackage{verbatim}
\usepackage{epstopdf}
\def\LB{\left(}
\def\RB{\right)}

\def\f{\frac}

\def\nn{\nonumber}
\def\R{\mathbf{R}}
\def\r{\mathbf{r}}

\def\k{\mathbf{k}}
\def\q{\mathbf{q}}

\def\bar{\begin{array}}
\def\ear{\end{array}}

\begin{document}

\title{Exact factorization-based density functional theory of electron-phonon systems}

\author{Ryan Requist}
\affiliation{Max Planck Institute of Microstructure Physics, Weinberg 2, 06120, Halle, Germany}

\author{C. R. Proetto}
\affiliation{Centro At\'omico Bariloche and Instituto Balseiro, 8400 San Carlos de Bariloche, R\'io Negro, Argentina}

\author{E. K. U. Gross}
\affiliation{Max Planck Institute of Microstructure Physics, Weinberg 2, 06120, Halle, Germany}
\affiliation{Fritz Haber Center for Molecular Dynamics, Institute of Chemistry, The Hebrew University of Jerusalem, Jerusalem 91904 Israel}

\date{\today}

\begin{abstract}
Density functional theory is generalized to incorporate electron-phonon coupling.  A Kohn-Sham equation yielding the electronic density $n_U(\r)$, a conditional probability density depending parametrically on the phonon normal mode amplitudes $U=\{U_{\q\lambda}\}$, is coupled to the nuclear Schr\"odinger equation of the exact factorization method.  The phonon modes are defined from the harmonic expansion of the nuclear Schr\"odinger equation.  A nonzero Berry curvature on nuclear configuration space affects the phonon modes, showing that the potential energy surface alone is generally not sufficient to define the phonons.  An orbital-dependent functional approximation for the non\-adiabatic exchange-correlation energy reproduces the leading-order nonadiabatic electron-phonon-induced band structure renormalization in the Fr\"ohlich model.
\end{abstract}

\maketitle

\section{\label{sec:introduction} Introduction}

The standard picture of interacting electrons and phonons in solids is a product of the Born-Oppenheimer (BO) approximation.\cite{born1927,born1954,ziman1960}  State-of-the-art first-principles calculations of electron-phonon-coupling effects start from a density functional theory (DFT) calculation for the equilibrium crystal structure.\cite{hohenberg1964,kohn1965}  The resulting ``clamped nuclei" electronic band structure depends on the BO approximation.  The normal modes of vibration (phonons) and first- and second-order electron-phonon coupling matrix elements are calculated from the response of the BO potential energy surface and Kohn-Sham (KS) orbitals to small displacements in the atomic positions.  The electronic band structure, phonon modes and electron-phonon coupling terms define a BO reference state that contains enough information to evaluate several observables, such as the electron-phonon coupling constant $\lambda$ and transition temperature $T_c$ in conventional superconductors such as MgB$_2$,\cite{kortus2001,kong2001,liu2001,bohnen2001,choi2002a,choi2002b,golubov2002,choi2003,mitrovic2004,choi2006,cappelluti2007,eiguren2008,choi2009a,calandra2010,delapenaseaman2010,margine2013,aperis2015} electronic band structure renormalization, \cite{allen1976,allen1981,*allen1981erratum,allen1983,kingsmith1989,eiguren2003,park2007,giustino2008,park2009,eiguren2009,giustino2010,cannuccia2011,gonze2011,antonius2015,ponce2015,*ponce2017erratum,antonius2016,nery2016,monserrat2016,allen2017} and electronic mass enhancement and specific heat.\cite{allen1972,golubov2002,choi2003,lin2008,delaire2008,choi2009a,choi2009b,subedi2009,diakhate2011,wei2013,tutuncu2013,tutuncu2015,zocco2015,li2015,wiendlocha2016,brown2016,zheng2017}  Nevertheless, there is growing interest in capturing nonadiabatic electron-phonon effects\cite{engelsberg1963,jarlborg1992,falter1995,maksimov1996,kulic2000,ferrari2007,basko2009,dean2010,defillipis2010,klimin2016,ponosov2016,ponosov2017} by {\it ab initio} approaches that go beyond this BO reference state.\cite{vanleeuwen2004,lazzeri2006,bock2006,pisana2007,piscanec2007,calandra2007,caudal2007,saitta2008,calandra2010,gonze2011,cannuccia2012,marini2015,ponce2015,antonius2015,dastuto2016,gali2016,nery2016,allen2017,giustino2017,long2017,zhou2017,caruso2017,nery2018,marini2018,novko2018,caruso2018}

Reliance on the BO approximation complicates subsequent many-body calculations.  Since the electronic Hamiltonian is already included in the adiabatic potential energy surface from which the reference BO phonons are calculated, it is not straightforward to rigorously divide the original electron-nuclear Hamiltonian into electronic $\hat{H}_{elec}$, phononic $\hat{H}_{phon}$, and electron-phonon coupling $\hat{H}_{elec-phon}$ terms,\cite{vanleeuwen2004} as typically done in setting up many-body perturbation theory.  It is therefore difficult to avoid double counting electronic interactions, and the many-body formalisms that have been proposed \cite{hedin1969,vanleeuwen2004,marini2015,antonius2015} are still more complicated than the widely-used BO-based approach outlined above, although work in this direction is ongoing.\cite{giustino2017,marini2018,karlsson2018}

To avoid double-counting issues, it would be desirable to be able to calculate electronic and phononic observables within a formally-exact DFT-like framework.  This is possible in multicomponent DFT,\cite{kreibich2001} where the functionals depend on both the electronic density $n(\r)$ in the body-fixed frame and the $N_n$-body nuclear density $\Gamma(\R_1,\R_2,\ldots,\R_{N_n})$, provided these densities can be realized in a noninteracting system with appropriate scalar potentials (noninteracting $v$ representablility).  A series of works \cite{lueders2005,marques2005,floris2005,floris2007} on superconducting DFT \cite{oliveira1988} have also been formulated to include $\Gamma(\R_1,\R_2,\ldots,\R_{N_n})$.  However, it has proven difficult to approximate the $\Gamma$-dependence of the exchange-correlation potentials in both multicomponent DFT and superconducting DFT.  Additionally, at temperature $T=0$, there does not exist an auxiliary noninteracting system capable of reproducing the density and anomalous density in superconducting DFT.\cite{schmidt2019}

In this paper, we focus on normal-state properties and show that a recent generalization of density functional theory \cite{requist2016b,li2018} based on the exact factorization (EF) of the electron-nuclear wavefunction into electronic and nuclear factors \cite{hunter1975,gidopoulos2014,abedi2010} offers a promising alternative for calculating electronic and phononic observables.  In contrast to multicomponent DFT's, the basic variable is a conditional electronic density $n_R(\r)$, a function which encodes the electronic density for each different set of nuclear coordinates $R=(\R_1,\R_2,\ldots)$.  Working with $n_R(\r)$ instead of the body-fixed-frame density $n(\r)$ makes the exchange-correlation functionals in this theory closer to those of standard BO-based DFT. 

Since EF-based DFT preserves the density-functional description of electronic structure that has made DFT so successful for solids, one can hope to obtain accurate approximations by building on the functionals of standard DFT.  Following Ref.~\onlinecite{li2018}, we consider an approximation strategy that consists in adding a nonadiabatic correction term to a standard DFT functional, such as a local density approximation (LDA) \cite{kohn1965,vonbarth1972} or a generalized gradient approximation (GGA).\cite{perdew1996}   Analytical calculations for the Fr\"ohlich model prove that this approximation achieves the correct leading-order electron-phonon-coupling induced band structure renormalization, including the velocity renormalization near the Fermi energy.

Section \ref{sec:theory} presents the general formalism of exact factorization-based density functional theory and its application to electron-phonon systems; Sec.~\ref{sec:phonons} defines phonons; Sec.~\ref{sec:functional} introduces our functional approximation; and Sec.~\ref{sec:Froehlich} applies the theory to the Fr\"ohlich model.  Conclusions and an outlook on future developments are given in Sec.~\ref{sec:conclusions}.

\section{\label{sec:theory} Exact factorization DFT}

\subsection{Electron-nuclear DFT \label{ssec:electron:nuclear:DFT}}

The exact factorization method \cite{hunter1975,gidopoulos2014,abedi2010} expresses the full electron-nuclear wavefunction as 
\begin{align}
\Psi(r,R) = \Phi_R(r) \chi(R) {,} \label{eq:factorization:rR}
\end{align}
where $r=\{\r_i\}$ denotes the set of electronic coordinates and $R=\{\R_{\mu}\}$ denotes the set of nuclear coordinates.   The key variable in exact factorization-based DFT\cite{requist2016b,li2018} is the conditional electronic density 
\begin{align}
n_R(\r) &= \frac{P(\r,R)}{P(R)} = \frac{N\int |\Psi(\r,\r_2,\ldots,\r_N,R)|^2 d\r_2 \ldots d\r_N}{\int |\Psi(\r_1,\r_2,\ldots,\r_N,R)|^2 d\r_1 \ldots d\r_N}\nn \\
&= \langle \Phi_R | \hat{\psi}^{\dag}(\r) \hat{\psi}(\r) | \Phi_R \rangle {,}
\end{align}
where $P(\r,R)$ is the joint probability to find an electron at position $\r$ and the nuclei at positions $R$ and $P(R)$ is the marginal probability of finding the nuclei at $R$ regardless of where the electrons are. 

The electronic density $n(\r)$ in a standard DFT calculation, which we hereafter denote as $n_R^{BO}(\r)$, is also a conditional density depending parametrically on $R$.  To see what beyond-BO contributions $n_R^{BO}(\r)$ is missing, consider the Born-Huang expansion \cite{born1954}
\begin{align}
\Psi(r,R) = \sum_{J=0}^{\infty} \Phi_{JR}^{BO}(r) \chi_J^{BO}(R) {,}
\end{align}
where $\Phi_{JR}^{BO}(r)$ is the $J^{th}$ eigenstate of the BO Hamiltonian
\begin{align}
\hat{H}^{BO} &= \sum_i \f{\mathbf{p}_i^2}{2m_e} + \sum_{\mu<\nu} \f{Z_{\mu} Z_{\nu} e^2}{4\pi\epsilon_0|\mathbf{R}_{\mu} - \mathbf{R}_{\nu}|} + \sum_{i<j} \f{e^2}{4\pi\epsilon_0|\mathbf{r}_i - \mathbf{r}_{j}|} \nn\\
&\quad - \sum_{i,\mu} \f{Z_{\mu} e^2}{4\pi\epsilon_0|\mathbf{r}_i - \mathbf{R}_{\mu}|} {.}
\end{align}
The exact conditional electronic density can be written in terms of the Born-Huang expansion as
\begin{align}
n_R(\mathbf{r}) = \frac{\sum_{J=0}^{\infty} |\chi_J^{BO}(R)|^2 \langle \Phi_{JR}^{BO} | \hat{\psi}^{\dag}(\r) \hat{\psi}(\r) | \Phi_{JR}^{BO}\rangle}{\sum_{J=0}^{\infty} |\chi_J^{BO}(R)|^2} {.}
\end{align}
A standard DFT calculation gives only the single term
\begin{align}
n_R^{BO}(\mathbf{r}) = \langle \Phi_{0R}^{BO} | \hat{\psi}^{\dag}(\r) \hat{\psi}(\r) | \Phi_{0R}^{BO}\rangle {.}
\end{align}

In terms of the nuclear wavefunction $\chi(R)$ and conditional electronic wavefunction $\Phi_R(r)$, the total energy of the electron-nuclear system can be expressed as
\begin{align}
E = T_{n,marg} + \int |\chi(R)|^2 \big[ \mathcal{E}^{BO}(R) + \mathcal{E}_{geo}(R) \big] dR {,} 
\label{eq:E:total}
\end{align}
where
\begin{align}
T_{n,marg} &= \int \chi^*(R) \sum_{\mu} \f{(\mathbf{P}_{\mu} + \mathbf{A}_{\mu})^2}{2M_{\mu}} \chi(R) dR {,} \nn \\
\mathcal{E}^{BO}(R) &= \langle \Phi_R | \hat{H}^{BO} | \Phi_R \rangle {,} \label{eq:Tnmarg} \\
\mathcal{E}_{geo}(R) &= \sum_{\mu} \f{\hbar^2 \big< \nabla_{\R_{\mu}} \Phi_R \big| \big(1-|\Phi_R\rangle \langle\Phi_R| \big) \big| \nabla_{\R_{\mu}} \Phi_R \big>}{2M_{\mu}} \nn
\end{align}
and $\mathbf{A}_{\mu}\equiv\hbar \mathrm{Im} \langle \Phi_R | \nabla_{\R_{\mu}} \Phi_R \rangle$.  Making the energy stationary with respect to variations of $\chi(R)$ and $\Phi_R(r)$ subject to the partial normalization condition $\int |\Phi_R(r)|^2 dr = 1$ for all $R$ leads to the following equations:\cite{gidopoulos2014}  
\begin{align}
\big[ \hat{H}^{BO} + \hat{U}_{en}\big] |\Phi_R\rangle &= \mathcal{E}(R) |\Phi_R\rangle \label{eq:electronic:R} \\
\left[ \sum_{\mu} \f{\big(\mathbf{P}_{\mu}+\mathbf{A}_{\mu}\big)^2}{2M_{\mu}} + \mathcal{E}(R) \right] \chi(R) &= E \chi(R) {,} \label{eq:nuclear:R} 
\end{align}
where $\hat{U}_{en}$ is a complicated operator that depends nonlinearly on $\chi$ and $\Phi_R$.  The nuclear equation has the form of a conventional Schr\"odinger equation with an exact potential energy surface $\mathcal{E} = \mathcal{E}^{BO}+\mathcal{E}_{geo}$ and an exact induced vector potential $\mathbf{A}_{\mu}$.

Exact factorization-based DFT\cite{requist2016b,li2018} seeks to bypass the many-body electronic equation, Eq.~(\ref{eq:electronic:R}), using in its place the conditional KS equation
\begin{align}
&\left[ \f{\mathbf{p}^2}{2m} + v_{en}(\r,R) + v_{hxc}^{EF}(\r,R)\right] \psi_{iR}(\r) = \epsilon_{iR} \psi_{iR}(\r) {,} \label{eq:KS:R} 
\end{align}
where $v_{en}(\r,R) = -\sum_{\mu}  Z_{\mu} e^2/4\pi\epsilon_0|\mathbf{r} - \mathbf{R}_{\mu}|$
and $v_{hxc}^{EF}(\r)$ is a nonadiabatic Hartree-exchange-correlation potential.

\subsection{Electron-phonon DFT}

We now consider a stable crystal with a well-defined equilibrium lattice structure.  Adopting notations similar to those in Refs.~\onlinecite{kwok1967,maradudin1968,giustino2017}, we specify the equilibrium position of nucleus $\kappa$ in primitive cell $l=(l_1,l_2,l_3)$ as 
\begin{align}
\R_{l\kappa}^{(0)} = \R_l^{(0)} + \boldsymbol{\tau}_{\kappa} {,}
\end{align}
where $\R_l^{(0)} = l_1 \mathbf{a}_1 + l_2 \mathbf{a}_2 + l_3 \mathbf{a}_3$ is the position of primitive cell $l$ and $\boldsymbol{\tau}_{\kappa}$ is the position of nucleus $\kappa$ within the primitive cell; $\mathbf{a}_i$ are the primitive lattice vectors.  The displacement of a nucleus from its equilibrium position is defined to be $\mathbf{u}_{l\kappa} =\R_{l\kappa}-\R_{l\kappa}^{(0)}$, and we denote the set of nuclear displacements as $u=\{\mathbf{u}_{l\kappa}\}$.

For electron-phonon systems, it is convenient to adopt Born-von K\'arm\'an boundary conditions and work with the phonon normal mode coordinates $U=\{U_{\q\lambda}\}$.  
Therefore, we introduce the factorization
\begin{align}
\Psi(r,U) = \Phi_U(r) \chi(U) {.} \label{eq:factorization:rU}
\end{align}
The phonon normal mode coordinates $U$ and their relationship to $u$ will be derived in the following section.  There are important differences with respect to standard DFT, where the relationship between $u$ and $U$ is
\begin{align}
\mathbf{u}_{l\kappa} &=  \sqrt{\f{M_0}{N M_{\kappa}}} \sum_{\q\lambda} U_{\q\lambda} \mathbf{e}_{\kappa}(\q\lambda) e^{i\q\cdot\R_l^{(0)}} \nn \\
\hat{U}_{\q\lambda} &= \sqrt{\f{M_{\kappa}}{N M_0}} \sum_{\kappa} \mathbf{u}_{l\kappa}\cdot \mathbf{e}_{\kappa}^*(\q\lambda) e^{-i\q\cdot\R_l^{(0)}} {.} \label{eq:transformation:0}
\end{align}
Here, $\mathbf{e}_{\kappa}(\q\lambda)$ is the polarization vector of the phonon normal mode, $M_0$ is an arbitrary reference mass, e.g.~the proton mass, and $N$ is the number of primitive cells under Born-von K\'arm\'an boundary conditions.  Throughout the paper, it is to be understood that the $\q=0$ acoustic modes 
%, which correspond to uniform translations of the crystal, 
are excluded from sums over the phonon quasimomentum.

The conditional KS equation in Eq.~(\ref{eq:KS:R}) becomes
\begin{align}
&\left[ \f{\mathbf{p}^2}{2m} + v_{en}(\r,u) + v_{hxc}^{EF}(\r,u)\right] \psi_{n \k u}(\r) = \epsilon_{n \k u} \psi_{n \k u}(\r) {,} \label{eq:KS} 
\end{align}
where $v_{en}(\r,u)  = -\sum_{l\kappa} Z_{\kappa} e^2/4\pi\epsilon_0|\r-\R_{l\kappa}^{(0)}-\mathbf{u}_{l\kappa}|$.
% and $v_{hxc}^{EF}(\r,u)$ will be derived in Sec.~\ref{sec:functional}.  
If we set the displacements $\mathbf{u}_{l\kappa}$ to zero, then the KS potential 
\begin{align}
v_s^{(0)}(\r) = \left.v_{en}(\r,u)\right|_{\mathbf{u}_{l\kappa}=0} + \left.v_{hxc}^{EF}(\r,u)\right|_{\mathbf{u}_{l\kappa}=0}
\end{align}
has lattice translational symmetry.  As in standard DFT, this allows us to label the KS orbitals with a band index $n$ and wavevector $\k$.  
In terms of the displacement coordinates $u$, Eq.~(\ref{eq:nuclear:R}) becomes
\begin{align}
\left[ \sum_{l\kappa} \f{\big(\mathbf{p}_{l\kappa}+\mathbf{A}_{l\kappa}(u)\big)^2}{2M_{\kappa}} + \mathcal{E}(u) \right] \chi(u) &= E \chi(u) {,} \label{eq:nuclear} 
\end{align}
where $\mathbf{p}_{l\kappa}=-i\hbar \nabla_{\mathbf{u}_{l\kappa}}$ and $\mathbf{A}_{l\kappa} = \hbar \mathrm{Im} \langle \Phi_u | \nabla_{\mathbf{u}_{l\kappa}} \Phi_u \rangle$.
Equations (\ref{eq:KS}) and (\ref{eq:nuclear}) are the fundamental equations of EF-based DFT for electron-phonon systems.  The exact potential energy surface $\mathcal{E}=\mathcal{E}^{BO}+\mathcal{E}_{geo}$ comprises a BO-like term 
\begin{align}
\mathcal{E}^{BO}(u) &= \langle \Phi_u | \hat{H}^{BO} |\Phi_u \rangle 
\end{align}
and a geometric term\cite{requist2016b,requist2016a}
\begin{align}
\mathcal{E}_{geo}(u) = \sum_{l\kappa} \f{\hbar^2 \big< \nabla_{\mathbf{u}_{l\kappa}} \Phi_u \big| \big(1-|\Phi_u\rangle \langle\Phi_u| \big) \big| \nabla_{\mathbf{u}_{l\kappa}} \Phi_u \big>}{2M_{\kappa}} {,}
\end{align}
which is similar to a term that can be derived in the BO approxima\-tion.\cite{berry1989,berry1990,berry1993}  $\mathcal{E}_{geo}$ is a geometric quantity that can be written as the contraction of a Riemannian metric tensor and an inverse mass tensor.

As a consequence of imposing Born-von K\'arm\'an boundary conditions, the exact potential energy surface, induced vector potential, and total energy in Eq.~(\ref{eq:nuclear}) acquire a parametric dependence on $\R_l^{(0)}$, i.e.,~on the lattice vectors $\mathbf{a}_i$.  
The equilibrium values of $\mathbf{a}_i$ can be obtained by minimizing the total energy at the end of the calculation.

Although separating off the center-of-mass motion, as we did in writing Eq.~(\ref{eq:factorization:rU}), modifies the electronic and nuclear kinetic energy operators,\cite{sutcliffe2000} the exact factorization scheme can still be straightforwardly applied to the resulting Schr\"odinger equation (see the supplemental material of Ref.~\onlinecite{requist2016b}).  
% ****** Transforming to a body-fixed frame induces rovibrational coupling, which, however, vanishes in the macroscopic limit relevant for solids.\cite{vanleeuwen2004}
To keep our focus on the essential differences between the present theory and standard DFT calculations of electron-phonon systems, we neglect these modifications and, moreover, we restrict our attention to nonpolar solids.

The induced vector potential $\mathbf{A}_{l\kappa}$ in Eq.~(\ref{eq:nuclear}) is said to be trivial if there exists a gauge choice such that $\mathbf{A}_{l\kappa}=0$.  This is not always the case.\cite{requist2016a,requist2017} In the following section, we show that the induced magnetic field (the curl of $\mathbf{A}_{l\kappa}$) affects the phonons.

\section{\label{sec:phonons} Exact phonons}

Phonons are usually calculated in the BO approximation.  The nuclear Schr\"odinger equation (\ref{eq:nuclear}) affords us a way of defining ``exact phonons." % , i.e.~without approximations.

We start by expanding $\mathcal{E}(u)$ and $\mathbf{A}_{l\kappa}(u)$ as
% in powers of the displacement from the equilibrium positions $R^{(0)} = \{\R_{l\kappa}^{(0)}\}$ as
\begin{align}
\mathcal{E}(u) &= \mathcal{E}_0 + \f{1}{2} \sum_{l\kappa\alpha l'\kappa'\alpha'} C_{l\kappa\alpha l'\kappa'\alpha'} u_{l\kappa\alpha} u_{l'\kappa'\alpha'} + \mathcal{O}(u^3) {,} \nn \\
A_{l\kappa\alpha}(u) &= A_{l\kappa\alpha}^{(0)} + \sum_{l'\kappa'\alpha'} A_{l\kappa\alpha l'\kappa'\alpha'}^{(1)} u_{l'\kappa'\alpha'} + \mathcal{O}(u^2) \label{eq:series} {,}
\end{align}
where $\alpha=(x,y,z)$, $A_{l\kappa\alpha,l'\kappa'\alpha'}^{(1)} = \partial A_{l\kappa\alpha}/\partial u_{l'\kappa'\alpha'}|_{u=0}$, and
\begin{align}
C_{l\kappa\alpha l'\kappa'\alpha'} = \f{\partial^2 \mathcal{E}}{\partial u_{l\kappa\alpha} \partial u_{l'\kappa'\alpha'}}\bigg|_{u=0} \label{eq:C}
\end{align}
is the force constant matrix.  We have assumed that the equilibrium coordinates $\R_{l\kappa}^{(0)}$ coincide with the coordinates $\R_{l\kappa}^{(min)}$ that minimize $\mathcal{E}(R)$.  This may not always be the case, particularly if the phononic wavefunction is delocalized on a strongly anharmonic potential energy surface.  The constant term $A_{l\kappa\alpha}^{(0)}$ in the expansion of $A_{l\kappa\alpha}$ can be removed by a gauge transformation $\Phi_u = \tilde{\Phi}_u e^{-(i/\hbar) A^{(0)}_{l\kappa\alpha} u_{l\kappa\alpha}}$ and is therefore inconsequential.  Thus, to second order, the Hamiltonian in Eq.~(\ref{eq:nuclear}) is 
\begin{align}
\hat{H}_{nucl}^{(2)} &= \sum_{l\kappa\alpha} \f{1}{2M_{\kappa}} \Big(\hat{p}_{l\kappa\alpha}+ \sum_{l'\kappa'\alpha'} A_{l\kappa\alpha l'\kappa'\alpha'}^{(1)} \hat{u}_{l'\kappa'\alpha'}\Big)^2 \nn \\
&+ \f{1}{2} \sum_{l\kappa\alpha l'\kappa'\alpha'} C_{l\kappa\alpha l'\kappa'\alpha'} \hat{u}_{l\kappa\alpha} \hat{u}_{l'\kappa'\alpha'} {.} \label{eq:Hn2:a}
\end{align}
Within the BO approximation, this form of Hamiltonian has been considered previously.\cite{holz1972,zhang2010} To find the eigenstates of $\hat{H}_{nucl}^{(2)}$, we first define the Fourier transformations 
\begin{align}
\hat{u}_{l\kappa\alpha} &= \sqrt{\f{M_0}{N M_{\kappa}}} \sum_{\q} \hat{u}_{\q\kappa\alpha} e^{i\q\cdot\R_l^{(0)}} \nn \\
\hat{u}_{\q\kappa\alpha} &= \sqrt{\f{M_{\kappa}}{N M_0}} \sum_{l} \hat{u}_{l\kappa\alpha} e^{-i\q\cdot\R_l^{(0)}} {.}
\end{align}
and  
\begin{align}
\hat{p}_{l\kappa\alpha} &= \sqrt{\f{M_{\kappa}}{N M_0}} \sum_{\q} \hat{p}_{\q\kappa\alpha} e^{i\q\cdot\R_l^{(0)}} \nn \\
\hat{p}_{\q\kappa\alpha} &= \sqrt{\f{M_0}{N M_{\kappa}}} \sum_{l} \hat{p}_{l\kappa\alpha} e^{-i\q\cdot\R_l^{(0)}} {.}
\end{align}
In the $\q$ representation, the Hamiltonian becomes
\begin{widetext}
\begin{align}
\hat{H}_{nucl}^{(2)} &= \f{1}{2M_0} \sum_{\q} \sum_{\kappa\alpha} \big[ \hat{p}_{-\q\kappa\alpha}  + \sum_{\kappa'\alpha'} A^{(1)}_{\kappa\alpha\kappa'\alpha'}(-\q) \hat{u}_{-\q\kappa'\alpha'} \big] \big[ \hat{p}_{\q\kappa\alpha} + \sum_{\kappa''\alpha''} A^{(1)}_{\kappa\alpha\kappa''\alpha''}(\q) \hat{u}_{\q\kappa''\alpha''} \big] \nn \\
&+\f{M_0}{2} \sum_{\q} \sum_{\kappa\alpha\kappa'\alpha'} \hat{u}_{-\q\kappa\alpha} D_{\kappa\alpha\kappa'\alpha'}(\q) \hat{u}_{\q\kappa'\alpha'} \label{eq:Hn2} {,}
\end{align}
\end{widetext}
where 
\begin{align}
A_{\kappa\alpha\kappa'\alpha'}^{(1)}(\q) &= \sum_{l'} \f{M_0}{\sqrt{M_{\kappa} M_{\kappa'}}} A_{0\kappa\alpha l'\kappa'\alpha'}^{(1)} e^{i\q\cdot\R_{l'}^{(0)}} \nn \\
D_{\kappa\alpha\kappa'\alpha'}(\q) &= \sum_{l'} \f{1}{\sqrt{M_{\kappa} M_{\kappa'}}} C_{0\kappa\alpha l'\kappa'\alpha'} e^{i\q\cdot\R_{l'}^{(0)}} {,} \label{eq:AD}
\end{align}
and the operators $\hat{u}_{\q\kappa\alpha}$ and $\hat{p}_{\q\kappa\alpha}$ satisfy the commutation relations
\begin{align}
[\hat{u}_{\q\kappa\alpha}, \hat{p}_{-\q'\kappa'\alpha'}] &= i\hbar \delta_{\q\q'} \delta_{\kappa\kappa'} \delta_{\alpha\alpha'} \nn \\
[\hat{u}_{\q\kappa\alpha}, \hat{u}_{\q'\kappa'\alpha'}] &= 0 \nn \\
[\hat{p}_{\q\kappa\alpha}, \hat{p}_{\q'\kappa'\alpha'}] &= 0 {.}
\end{align}
To diagonalize the Hamiltonian in Eq.~(\ref{eq:Hn2}), we first apply the canonical transformation 
\begin{align}
\hat{u}_{\q\kappa\alpha} &= \sum_{\lambda} \big[ e_{\kappa\alpha}(\q\lambda) \hat{U}_{\q\lambda} + d_{\kappa\alpha}(\q\lambda) \hat{P}_{\q\lambda}\big] \nn \\
\hat{p}_{\q\kappa\alpha} &= \sum_{\lambda} \big[ -d_{\kappa\alpha}(\q\lambda) \hat{U}_{\q\lambda} + e_{\kappa\alpha}(\q\lambda) \hat{P}_{\q\lambda}\big] {.} \label{eq:trans}
\end{align}
The inverse transformation is
\begin{align}
\hat{U}_{\q\lambda} &= \sum_{\kappa\alpha} \big[ e_{\kappa\alpha}^*(\q\lambda) \hat{u}_{\q\kappa\alpha} - d_{\kappa\alpha}^*(\q\lambda) \hat{p}_{\q\kappa\alpha} \big] \nn \\
\hat{P}_{\q\lambda} &= \sum_{\kappa\alpha} \big[ d_{\kappa\alpha}^*(\q\lambda) \hat{u}_{\q\kappa\alpha} + e_{\kappa\alpha}^*(\q\lambda) \hat{p}_{\q\kappa\alpha} \big] \label{eq:trans:inv} {.}
\end{align}
$\hat{U}_{\q\lambda}$ and $\hat{P}_{\q\lambda}$ should satisfy the commutation relations 
\begin{align}
[\hat{U}_{\q\lambda},\hat{P}_{-\q'\lambda'}] &= i\hbar \delta_{\q\q'} \delta_{\lambda\lambda'} \nn \\
[\hat{U}_{\q\lambda},\hat{U}_{\q'\lambda'}] &= 0 \nn \\
[\hat{P}_{\q\lambda},\hat{P}_{\q'\lambda'}] &= 0 {.}
\end{align}
The transformation in Eqs.~(\ref{eq:trans}) and (\ref{eq:trans:inv}) will preserve the commutation relations if the polarization vectors $e_{\kappa\alpha}(\q\lambda)$ and $d_{\kappa\alpha}(\q\lambda)$ satisfy the orthonormality conditions
\begin{align}
\sum_{\kappa\alpha} \big[ e_{\kappa\alpha}^*(\q\lambda) e_{\kappa\alpha}(\q\lambda') + d_{\kappa\alpha}^*(\q\lambda) d_{\kappa\alpha}(\q\lambda') \big] &= \delta_{\lambda\lambda'} \nn \\
\sum_{\lambda} \big[ e_{\kappa\alpha}(\q\lambda) e_{\kappa'\alpha'}^*(\q\lambda) + d_{\kappa\alpha}(\q\lambda) d_{\kappa'\alpha'}^*(\q\lambda) \big] &= \delta_{\kappa\kappa'} \delta_{\alpha\alpha'} \label{eq:normalization} 
\end{align}
and
\begin{align}
\sum_{\kappa\alpha} \big[ e_{\kappa\alpha}^*(\q\lambda) d_{\kappa\alpha}(\q\lambda') - d_{\kappa\alpha}^*(\q\lambda) e_{\kappa\alpha}(\q\lambda') \big] &= 0 \nn \\
\sum_{\lambda} \big[ e_{\kappa\alpha}(\q\lambda) d_{\kappa'\alpha'}^*(\q\lambda) - d_{\kappa\alpha}(\q\lambda) e_{\kappa'\alpha'}^*(\q\lambda) \big] &= 0 \label{eq:orthogonality} {.}
\end{align}
The transformation can be summarized with the help of a matrix $V$ as
\begin{align}
\LB \bar{c} [\hat{u}_{\q}] \\ \hspace{0cm} [\hat{p}_{\q}] \ear \RB = \underbrace{\LB \begin{array}{rr} [e(\q)] & [d(\q)] \\ -[d(\q)] & [e(\q)] \end{array} \RB}_{V} \LB \begin{array}{c} [\hat{U}_{\q}] \\ \hspace{0cm}[\hat{P}_{\q}] \end{array} \RB {,}
\end{align}
where $[\hat{u}_{\q}]$, $[\hat{p}_{\q}]$, $[\hat{U}_{\q}]$, $[\hat{P}_{\q}]$, $[e(\q)]$ and $[d(\q)]$ are columns and matrices indexed by $\kappa\alpha$ and $\lambda$.  Then, the orthonormality constraints can be succinctly expressed as % $V^{\dag} V = I$ and $V V^{\dag} = I$.
\begin{align}
V^{\dag} V &= \LB \bar{cc} [\delta_{\lambda\lambda'}] & 0 \\ 0 & [\delta_{\lambda\lambda'}] \ear \RB \nn \\
V V^{\dag} &= \LB \bar{cc} [\delta_{\kappa\kappa'} \delta_{\alpha\alpha'}] & 0 \\ 0 & [\delta_{\kappa\kappa'} \delta_{\alpha\alpha'}] \ear \RB {.}
\end{align}
%\begin{align}
%\LB \begin{array}{c} [\hat{u}_{\q}]_{\kappa\alpha} \\ [\hat{p}_{\q}]_{\kappa\alpha} \end{array} \RB = \LB \begin{array}{rr} [e(\q)]_{\kappa\alpha,\lambda} & [d(\q)]_{\kappa\alpha,\lambda} \\ -[d(\q)]_{\kappa\alpha,\lambda} & [e(\q)]_{\kappa\alpha,\lambda} \end{array} \RB \LB \begin{array}{c} [\hat{U}_{\q}]_{\lambda} \\ [\hat{P}_{\q}]_{\lambda} \end{array} \RB {.}
%\end{align}

Applying the above transformation to Eq.~(\ref{eq:Hn2}) and requiring the coefficients of the $\hat{U}_{-\q\lambda} \hat{P}_{\q\lambda}$ and $\hat{P}_{-\q\lambda} \hat{U}_{\q\lambda}$ terms to vanish leads to the following eigenvalue equations:
\begin{align}
\f{1}{M_0} \left( \begin{array}{cc} I & A^{(1)} \\ A^{(1)\dag} & M_0^2 D + A^{(1)\dag} A^{(1)} \end{array} \right) \left( \begin{array}{c} e \\ d \end{array} \right) &= \f{1}{\mathcal{M}} \left( \begin{array}{c} e \\ d \end{array} \right) \nn \\
\f{1}{M_0} \left( \begin{array}{cc} I & A^{(1)} \\ A^{(1)\dag} & M_0^2 D + A^{(1)\dag} A^{(1)} \end{array} \right) \left( \begin{array}{c} -d \\ e \end{array} \right) &= \mathcal{M} \Omega^2 \left( \begin{array}{c} -d \\ e \end{array} \right) {,} \label{eq:eigenvalue}
\end{align}
where $A^{(1)}$ and $D$ are the matrices in Eq.~(\ref{eq:AD}), $I$ is the identity matrix, and $1/\mathcal{M}$ and $\mathcal{M}\Omega^2$ are eigenvalues; $\q\lambda$ has been suppressed.  The two types of eigenvectors in Eq.~(\ref{eq:eigenvalue}) are 
\begin{align}
v_P(\q\lambda) &= \left( \begin{array}{c} e_{\alpha_1\kappa_1}(\q\lambda) \\ e_{\alpha_2\kappa_2}(\q\lambda) \\ \vdots \\ d_{\alpha_1\kappa_1}(\q\lambda) \\ d_{\alpha_2\kappa_2}(\q\lambda) \\ \vdots \end{array} \right) \quad
v_U(\q\lambda) = \left( \begin{array}{c} -d_{\alpha_1\kappa_1}(\q\lambda) \\ -d_{\alpha_2\kappa_2}(\q\lambda) \\ \vdots \\ e_{\alpha_1\kappa_1}(\q\lambda) \\ e_{\alpha_2\kappa_2}(\q\lambda) \\ \vdots \end{array} \right) {.}
\end{align}
The identities $D_{\kappa\alpha\kappa'\alpha'}(-\q)=D_{\kappa\alpha\kappa'\alpha'}^*(\q)$ and $A_{\kappa\alpha\kappa'\alpha'}^{(1)}(-\q)=A_{\kappa\alpha\kappa'\alpha'}^{(1)*}(\q)$, which follow from the definitions in Eq.~(\ref{eq:AD}), together with $e_{\kappa\alpha}(-\q)=e_{\kappa\alpha}^*(\q)$ and $d_{\kappa\alpha}(-\q)=d_{\kappa\alpha}^*(\q)$, imply that if $v_P(\q\lambda)$ is an eigenvector with eigenvalue $1/\mathcal{M}_{\q\lambda}$, then $v_P(-\q\lambda)$ is also an eigenvector with the same eigenvalue, i.e.~$\mathcal{M}_{\q\lambda}=\mathcal{M}_{-\q\lambda}$. Similar considerations for $v_U(\q\lambda)$ imply $\Omega_{\q\lambda}=\Omega_{-\q\lambda}$.  Equation~(\ref{eq:eigenvalue}) replaces the standard eigenvalue equation $D e = \omega^2 e$ defining the phonons in terms of the dynamical matrix $D$.  In Sec.~\ref{ssec:force:constant:matrix}, we discuss the differences between phonon calculations in our theory and standard calculations in density functional perturbation theory.\cite{baroni2001,savrasov1996,gonze1997a}  
% Eqs.~(\ref{eq:Hn2:a}) and (\ref{eq:eigenvalue})

After these preliminaries, the transformed Hamiltonian can be expressed as
\begin{align}
\hat{H}_{nucl}^{(2)} = \sum_{\q\lambda} \bigg[ \f{1}{2\mathcal{M}_{\q\lambda}} \hat{P}_{-\q\lambda} \hat{P}_{\q\lambda} + \f{\mathcal{M}_{\q\lambda}\Omega_{\q\lambda}^2}{2}  \hat{U}_{-\q\lambda} \hat{U}_{\q\lambda} \bigg] {,} \label{eq:H:PU}
\end{align}
where $\mathcal{M}_{\q\lambda}$ enters as a $\q\lambda$-dependent effective mass.  Finally, in terms of the creation and annihilation operators
\begin{align}
a_{\q\lambda}^{\dag} &= \sqrt{\f{\mathcal{M}_{\q\lambda}\Omega_{\q\lambda}}{2\hbar}} \hat{U}_{-\q\lambda} - \f{i}{\sqrt{2\hbar \mathcal{M}_{\q\lambda} \Omega_{\q\lambda}}} \hat{P}_{\q\lambda} \nn \\
a_{\q\lambda} &= \sqrt{\f{\mathcal{M}_{\q\lambda}\Omega_{\q\lambda}}{2\hbar}} \hat{U}_{\q\lambda} + \f{i}{\sqrt{2\hbar \mathcal{M}_{\q\lambda} \Omega_{\q\lambda}}} \hat{P}_{-\q\lambda} {,}
\end{align} 
the harmonic phonon Hamiltonian becomes
\begin{align}
\hat{H}_{nucl}^{(2)} = \sum_{\q\lambda} \hbar\Omega_{\q\lambda} \Big( a_{\q\lambda}^{\dag} a_{\q\lambda} + \f{1}{2} \Big) {.} \label{eq:Hnucl}
\end{align}
Since this is bilinear in $a_{\q\lambda}^{\dag}$ and $a_{\q\lambda}$, there are no phonon interactions at this order.  It was for the purpose of obtaining this result that the expansion of $A_{l\kappa\alpha}$ was terminated at the first order; the second-order terms would have generated terms in Eq.~(\ref{eq:Hnucl}) that are cubic and quartic in $a_{\q\lambda}^{\dag}$ and $a_{\q\lambda}$.

The occurrence of a nonvanishing Berry curvature $B_{l\kappa\alpha,l'\kappa'\alpha'}=A^{(1)}_{l'\kappa'\alpha'l\kappa\alpha} - A^{(1)}_{l\kappa\alpha,l'\kappa'\alpha'}$ on nuclear configuration space implies time-reversal symmetry breaking; this occurs naturally if the electronic state breaks time-reversal symmetry, e.g.~in magnetic or (anomalous) quantum Hall systems.  Our analysis is similar to that of Ref.~\onlinecite{holz1972}, where external rather than induced magnetic fields were considered.  

In the special case $A^{(1)}=0$, $\mathcal{M}_{\q\lambda} \rightarrow M_0$, $\Omega_{\q\lambda}^2 \rightarrow \omega_{\q\lambda}^2$, and Eq.~(\ref{eq:eigenvalue}) reduces to a single eigenvalue equation
\begin{align}
D e = \omega^2 e {,}
\end{align}
and the transformation in Eqs.~(\ref{eq:trans}) and (\ref{eq:trans:inv}) reduces to Eq.~(\ref{eq:transformation:0}). Hence, in this case Eq.~(\ref{eq:H:PU}) recovers the standard Hamiltonian
\begin{align}
\hat{H}_{nucl}^{(2)} = \sum_{\q\lambda} \bigg[ \f{1}{2M_0} \hat{P}_{-\q\lambda} \hat{P}_{\q\lambda} + \f{M_0}{2} \omega_{\q\lambda}^2 \hat{U}_{-\q\lambda} \hat{U}_{\q\lambda} \bigg] {.}
\end{align}
It will be helpful to write the explicit harmonic ground state wavefunction in the $U$ representation.  Using the orthogonality of the phonon modes, it is
\begin{align}
\chi_0(U) = \langle U | \chi_0 \rangle = \prod_{\q\lambda} \LB \f{M_0\omega_{\q\lambda}}{\pi\hbar} \RB^{\f{1}{4}} \exp\LB -\f{U_{\q\lambda} U_{-\q\lambda}}{4L_{\q\lambda}^2}\RB {,}
\label{eq:chi0}
\end{align}
where $L_{\q\lambda}$ is the amplitude of zero-point motion 
\begin{align}
L_{\q\lambda} = \sqrt{\f{\hbar}{2M_0\omega_{\q\lambda}}} {.}
\end{align}

Phonon-phonon interactions arise from the anharmonicity of the potential energy surface and higher-order terms in the expansion of $A_{l\kappa\alpha}$, e.g.~at third-order 
\begin{align*}
\hat{H}_{nucl}^{(3)} = \sum_{\q_1\lambda_1,\q_2\lambda_2,\q_3\lambda_3} \Gamma_{\q_1\lambda_1,\q_2\lambda_2,\q_3\lambda_3} (a_{\q_1\lambda_1}+a_{-\q_1\lambda_1}^{\dag}) \nn \\
\times (a_{\q_2\lambda_2}+a_{-\q_2\lambda_2}^{\dag}) (a_{\q_3\lambda_3}+a_{-\q_3\lambda_3}^{\dag}) \delta_{\q_1+\q_2+\q_3,\mathbf{G}}  {.} 
\end{align*}
The full Schr\"odinger equation for the phonons then has the form
\begin{align}
\left[ \hat{H}_{nucl}^{(2)} + \hat{H}_{nucl}^{(3)} + \cdots \right] |\chi\rangle = E |\chi\rangle {,} \label{eq:chi}
\end{align}
and, in practice, the phonon-phonon interactions must be truncated at some order.  

\section{\label{sec:functional} Nonadiabatic Hartree-exchange-correlation functional}

From now on, we assume that the induced vector potential $\mathbf{A}_{l\kappa}$ in Eq.~(\ref{eq:nuclear}) is trivial; this assumption can be relaxed.  Using the transformation [Eq.~(\ref{eq:transformation:0})] from nuclear displacements $u=\{\mathbf{u}_{l\kappa}\}$ to phonon normal mode coordinates $U=\{U_{\q\lambda}\}$, Eqs.~(\ref{eq:KS}) and (\ref{eq:nuclear}) become
\begin{align}
\left[ \f{\mathbf{p}^2}{2m} + v_{en}(\r,U) + v_{hxc}^{EF}(\r,U)\right]& \psi_{n \k U}(\r) = \epsilon_{n \k U} \psi_{n \k U}(\r) {,} \label{eq:KS:U} \\
\bigg[ \sum_{\q\lambda} \f{\hat{P}_{-\q\lambda}\hat{P}_{\q\lambda}}{2M_{0}} + \mathcal{E}(U) \bigg] &\chi(U) = E \chi(U) {,} \label{eq:nuclear:U} 
\end{align}
where $\hat{P}_{\q\lambda}=-i\hbar\partial/\partial U_{-\q\lambda}$, $\mathcal{E}(U)=\mathcal{E}^{BO}(U)+\mathcal{E}_{geo}(U)$, 
\begin{align}
\mathcal{E}^{BO}(U) &= \langle \Phi_U | \hat{H}^{BO} | \Phi_U\rangle {,} \nn \\
\mathcal{E}_{geo}(U) &= \f{\hbar^2}{2M_0} \sum_{\q\lambda} \bigg< \f{\partial \Phi_U}{\partial U_{\q\lambda}} \bigg| \Big( 1- \Big| \Phi_U \Big> \Big< \Phi_U \Big| \bigg) \bigg| \f{\partial \Phi_U}{\partial U_{\q\lambda}} \bigg> {,}\label{eq:Egeo:2}
\end{align}
and we recall that $\q=0$ acoustic modes are omitted from all sums.

The total energy of the electron-phonon system is
\begin{align}
E &= \int \chi^*(U) \bigg[ \sum_{\q\lambda} \f{\hat{P}_{-\q\lambda}\hat{P}_{\q\lambda}}{2M_0} + \mathcal{E}(U) \bigg] \chi(U) dU {.} \label{eq:E}
\end{align}
One of the advantages of unifying electrons and phonons in a DFT framework is that a single density functional approximation for $\mathcal{E}(U)$ determines, on equal footing, all of the potentials in Eqs.~(\ref{eq:KS:U}) and (\ref{eq:nuclear:U}).  

As in standard DFT, the conditional electronic density is obtained from the occupied orbitals according to
\begin{align}
n_U(\r) &= \sum_{n \k} f_{n\k U} |\psi_{n \k U}(\r)|^2 {,}
\label{eq:density}
\end{align}
where $f_{n\k U}$ is a $U$-dependent occupation number.  From now on, we suppress the subscript $U$ on the occupation numbers and orbitals.

Reference~\onlinecite{requist2016b} introduced an exact factorization-based DFT in which the energy is expressed as a variational functional of $(n_R,\mathbf{j}_{pR},A_{\mu},\mathcal{T}_{\mu\nu},\chi)$, where $\mathbf{j}_{pR}$ is the conditional electronic paramagnetic current density, $A_{\mu}$ is the induced vector potential and $\mathcal{T}_{\mu\nu}$ is the quantum geometric tensor.\cite{berry1989,provost1980}  Reference~\onlinecite{li2018} showed that the energy can also be expressed as a functional of $(n_R,|\chi|^2)$.  We take a similar approach here and interpret the energy of an electron-phonon system as a functional of $(n_U,|\chi|^2)$.  Equations~(\ref{eq:KS:U}) and (\ref{eq:nuclear:U}) are then coupled through the functional dependence of the potentials: $\mathcal{E}$ depends on the density $n_U$, while $v_{hxc}^{EF}$ depends on $|\chi|^2$ (and $n_U$).

\subsection{Approximation strategy}

Our strategy for approximating $\mathcal{E}[n_U]$ is the following.  First, noting that $\mathcal{E}^{BO}[n_U]$ can be written as in DFT as
\begin{align}
\mathcal{E}^{BO}[n_U]&=T_s[n_U]+\int V_{en}(\r,U) n_U(\r) d\r  + V_{nn}(U) \nn \\
&+E_{hxc}[n_U] {,}
\end{align}
we approximate $E_{hxc}[n_U]$ by a standard semilocal BO-based DFT functional $E_{hxc}^{BO}[n_U]$ such as a GGA.  Second, in the nonadiabatic term $\mathcal{E}_{geo}$ in Eq.~(\ref{eq:Egeo:2}), we approximate the correlated electronic wavefunction $|\Phi_U\rangle$ by the Slater determinant of occupied KS orbitals.  This defines an orbital-dependent functional $\mathcal{E}_{geo}[\psi_{n\k}]$, which is only implicitly a functional of $n_U$.  The essential feature of this approximation is that nuclear mass-dependent, nonadiabatic effects are described by a simple additive correction to an existing DFT functional.  

With this approximation for $\mathcal{E}_{geo}$, we have 
\begin{widetext}
\begin{align}
\mathcal{E}_{geo}[\psi_{n\k}] &= \f{\hbar^2}{2M_0} \sum_{\q\lambda} \sum_{n\k} f_{n\k} \bigg< \f{\partial \psi_{n\k}}{\partial U_{\q\lambda}} \bigg| \Big( 1 - \Big| \psi_{n\k} \Big> \Big<\psi_{n\k} \Big| \Big) \bigg| \f{\partial \psi_{n\k}}{\partial U_{\q\lambda}} \bigg> - \f{\hbar^2}{2M_0} \sum_{\q\lambda} \sum_{n\k \neq n^{\prime}\k^{\prime}} f_{n\k} f_{n'\k'} \bigg< \f{\partial \psi_{n\k}}{\partial U_{\q\lambda}} \bigg| \psi_{n^{\prime}\k^{\prime}} \bigg> \bigg< \psi_{n^{\prime}\k^{\prime}} \bigg|  \f{\partial \psi_{n\k}}{\partial U_{\q\lambda}} \bigg> \label{eq:Egeo:approx} {.}
\end{align}
\end{widetext}
Via a chain rule for orbital-dependent functionals,\cite{kuemmel2008} e.g.
\begin{align}
v_{geo}(\r,U) &= \sum_{n\k} \iint \f{\delta \mathcal{E}_{geo}}{\delta \psi_{n\k}(\r'')} \f{\delta \psi_{n\k}(\r'')}{\delta v_s(\r')} \f{\delta v_s(\r')}{\delta n_U(\r)} d\r' d\r''  \nn \\
&+ c.c. + \cdots
\end{align}
the above approximations yield the scalar nonadiabatic Hartree-exchange-correlation potential in Eq.~(\ref{eq:KS:U}), i.e.
\begin{align}
v_{hxc}^{EF}(\r,U) = v_{hxc}^{BO}(\r,U) + v_{geo}(\r,U) {,}
\end{align} 
where $v_{hxc}^{BO}(\r,U)= \delta E_{hxc}^{BO}/\delta n_U(\r)$ is the standard DFT potential and $v_{geo}(\r,U)$ is a nonadiabatic correction. 

Our analytical calculations for the Fr\"ohlich model in Sec.~\ref{sec:Froehlich} suggest that using a nonlocal (orbital-dependent) exchange-correlation potential is a more natural way to incorporate electron-phonon coupling.  Since $\mathcal{E}^{BO}[n_U]$ can be converted into an orbital-dependent functional by substituting $n_U=\sum_{n \k} f_{n\k} |\psi_{n \k}|^2$, our approximation $\mathcal{E}_{geo}[\psi_{n\k}]$ implies an approximation for $\mathcal{E}[\psi_{n\k}] = \mathcal{E}^{BO}[\psi_{n\k}] + \mathcal{E}_{geo}[\psi_{n\k}]$ and, in turn, the total energy in Eq.~(\ref{eq:E}). The stationary conditions with respect to $\psi_{n\k}^*(\r)$ lead to a nonadiabatic and nonlocal generalized KS potential of the form:
\begin{align}
\hat{v}_s(U)  = \hat{v}_{en}(U) + \hat{v}_{hxc}^{\rm BO}(U) + \hat{v}_{geo}(U) {,}
\label{eq:vs:operator}
\end{align}
where $\hat{v}_{en}(U)$ and $\hat{v}_{hxc}^{\rm BO}(U)$ are the usual local potentials $v_{en}(\r,U)$ and $v_{hxc}^{\rm BO}(\r,U)$, and $\hat{v}_{geo}(U)$ is defined by its matrix elements
\begin{align}
&\langle \psi_{m\k+\q} | \hat{v}_{geo}| \psi_{n\k}\rangle = \nn\\
&\quad - \f{1}{|\chi|^2} \sum_{\lambda} \bigg< \psi_{m\k+\q} \bigg| \f{\partial}{\partial U_{\q\lambda}^*} \bigg[ |\chi|^2  \f{\delta \mathcal{E}_{geo}}{\delta (\partial \psi_{n\k}^*/\partial U_{\q\lambda}^*)} \bigg] \bigg> \nn \\
&=-\f{\hbar^2}{2M_0} \sum_{\lambda} \f{\partial \ln|\chi|^2}{\partial U_{-\q\lambda}} \bigg< \psi_{m\k+\q} \bigg| \f{\partial\psi_{n\k}}{\partial U_{\q\lambda}} \bigg> \nn \\
&\quad -\f{\hbar^2}{2M_0} \sum_{\lambda} \bigg< \psi_{m\k+\q} \bigg| \f{\partial^2\psi_{n\k}}{\partial U_{-\q\lambda} \partial U_{\q\lambda}} \bigg> {.}
\end{align}
The first term, hereafter denoted as $\langle \psi_{m\k+\q} | \hat{v}_{geo}^{(1)} | \psi_{n\k}\rangle$, is first order in $g$. The second term is second order.  For the nuclear wavefunction in Eq.~(\ref{eq:chi0}), we have
\begin{align}
\f{\partial \ln |\chi_0|^2}{\partial U_{-\q\lambda}} = - \f{U_{\q\lambda}}{L_{\q\lambda}^2} {,}
\end{align}
so that to leading order we can write 
\begin{align}
\langle \psi_{m\k+\q} | \hat{v}_{geo} | \psi_{n\k}\rangle &\approx
\langle \psi_{m\k+\q} | \hat{v}_{geo}^{(1)} | \psi_{n\k}\rangle \nn \\
&= \sum_{\lambda} \f{\hbar^2}{2M_0} \f{U_{\q\lambda}}{L_{\q\lambda}^2} \bigg< \psi_{m\k+\q} \bigg| \f{\partial\psi_{n\k}}{\partial U_{\q\lambda}} \bigg> \nn \\
&= \sum_{\lambda} \hbar \omega_{\q\lambda} U_{\q\lambda} \bigg< \psi_{m\k+\q} \bigg| \f{\partial\psi_{n\k}}{\partial U_{\q\lambda}} \bigg> {.} \label{eq:vgeo:offdiag}
\end{align}
The factor $\langle \psi_{m\k+\q} | \partial\psi_{n\k}/\partial U_{\q\lambda} \rangle$ would in practice be determined self-consistently during the solution of Eqs.~(\ref{eq:KS:U}) and (\ref{eq:nuclear:U}).   Similarly, from the stationary condition with respect to variations of $f_{n\k}$, we obtain a second-order diagonal contribution
\begin{align}
\langle \psi_{n\k} | \hat{v}_{geo}^{(2)} | \psi_{n\k}\rangle 
&= \sum_{\lambda} \f{\hbar^2}{2M_0} \bigg< \psi_{n\k} \bigg| \f{\partial^2 \psi_{n\k}}{\partial U_{-\q\lambda} \partial U_{\q\lambda}} \bigg> {.} \label{eq:vgeo:diag}
\end{align}
Equations~(\ref{eq:KS:U}) and (\ref{eq:nuclear:U}) together with $E_{hxc}^{BO}[n_U]$ and Eqs.~(\ref{eq:Egeo:approx}), (\ref{eq:vgeo:offdiag}) and (\ref{eq:vgeo:diag}) completely determine the exact factorization DFT equations to second order in $g$.  In Sec.~\ref{sec:Froehlich}, we apply these equations to the Fr\"ohlich model and demonstrate that they exactly recover the leading-order nonadiabatic electron-phonon coupling effects.

The operator $\hat{v}_s(U)$ in Eq.~(\ref{eq:vs:operator}) is not a scalar multiplicative potential and therefore takes us outside a strict KS framework.  As a result, the single-particle orbitals $\psi_{n\k}(\r)$ will not generally equal the KS orbitals.\cite{kuemmel2008}

\subsection{\label{ssec:force:constant:matrix} Evaluating the force constant matrix}

Given any approximation for $\mathcal{E}(u)$, we can evaluate the force constant matrix $C_{l\kappa\alpha l'\kappa'\alpha'}=\partial^2 \mathcal{E}/\partial u_{l\kappa\alpha} \partial u_{l'\kappa'\alpha'}$.  By virtue of the transformation in Eq.~(\ref{eq:transformation:0}), our approximation $\mathcal{E}(U) = T_s[n_U]+\int V_{en}(\r,U) n_U(\r) d\r  + V_{nn}(U) +E_{hxc}^{BO}[n_U] + \mathcal{E}_{geo}[\psi_{n\k U}]$ implies an approximation for $\mathcal{E}(u)$.  Since the $u$ dependence enters both explicitly, through $V_{en}$ and $V_{nn}$, and implicitly, through the functional dependence on $n_u$ and $\psi_{n\k u}$, we obtain
\begin{widetext}
\begin{align}
\f{\partial^2 \mathcal{E}}{\partial u_{l\kappa\alpha} \partial u_{l'\kappa'\alpha'}} &= \f{\partial^2 V_{nn}}{\partial u_{l\kappa\alpha} \partial u_{l'\kappa'\alpha'}} + \int \f{\partial^2 V_{en}}{\partial u_{l\kappa\alpha} \partial u_{l'\kappa'\alpha'}} n_u(\r) d\r + \int \bigg( \f{\partial V_{en}}{\partial u_{l\kappa\alpha}} \f{\partial n_u(\r)}{\partial u_{l'\kappa'\alpha'}} + \f{\partial V_{en}}{\partial u_{l'\kappa'\alpha'}} \f{\partial n_u(\r)}{\partial u_{l\kappa\alpha}} \bigg) d\r \nn \\
&+ \int V_{en} \f{\partial^2 n_u(\r)}{\partial u_{l\kappa\alpha} \partial u_{l'\kappa'\alpha'}} d\r + \iint \f{\delta^2 (T_s+ E_{hxc}^{BO})}{\delta n_u(\r) \delta n_u(\r')} \f{\partial n_u(\r)}{\partial u_{l\kappa\alpha}} \f{\partial n_u(\r')}{\partial u_{l'\kappa'\alpha'}} d\r d\r' + \int \f{\delta (T_s+ E_{hxc}^{BO})}{\delta n_u(\r)} \f{\partial^2 n_u(\r)}{\partial u_{l\kappa\alpha} \partial u_{l'\kappa'\alpha'}} d\r \nn \\ 
&+ \f{\partial^2 \mathcal{E}_{geo}}{\partial u_{l\kappa\alpha} \partial u_{l'\kappa'\alpha'}} {.} \label{eq:Hessian}
\end{align}
\end{widetext}
The last term generally leads to many terms involving the chain rule, e.g.~
\begin{align}
\sum_{n\k,l''\kappa''\alpha''} \int \f{\delta \mathcal{E}_{geo}}{\delta (\partial \psi_{n\k}(\r)/\partial u_{l''\kappa''\alpha''})} \f{\partial (\partial \psi_{n\k}(\r)/\partial u_{l''\kappa''\alpha''})}{\partial u_{l\kappa\alpha}} d\r {.}
\end{align}

The first-order density response $\partial n_u(\r)/\partial u_{l\kappa\alpha}$ plays an essential role in Eq.~(\ref{eq:Hessian}), just as it does in density functional perturbation theory (DFPT)\cite{baroni2001,savrasov1996,gonze1997a} [c.f.~Eq.~(10) in Ref.~\onlinecite{baroni2001}]. There are a few important distinctions between phonon calculations in EF-based DFPT and standard BO-based DFPT.  First, we cannot use the Hellmann-Feynman theorem, since $\mathbf{u}_{l\kappa}$ are not merely parameters in the exact electronic Schr\"odinger equation [the operator $\hat{U}_{en}$ in Eq.~(\ref{eq:electronic:R}) contains the gradient $\nabla_{\mathbf{R}_{l\kappa}} = \nabla_{\mathbf{u}_{l\kappa}}$].  As a result, the Hessian of $\mathcal{E}$ also depends on the second-order density response $\partial^2 n_u(\r)/\partial u_{l\kappa\alpha} \partial u_{l'\kappa'\alpha'}$.  Second, our theory includes an induced vector potential $\mathbf{A}_{l\kappa}$ in the exact nuclear Schr\"odinger equation, which, if nontrivial, affects the phonon modes, showing that the force constant matrix alone is generally not sufficient to define the exact phonons.  Through the self-consistent solution of Eqs.~(\ref{eq:KS:U}) and (\ref{eq:nuclear:U}), we achieve a nonadiabatic extension of standard DFPT. Only marginally more computational time and resources are needed for an EF-based DFPT calculation than for a standard DFPT calculation.

\section{\label{sec:Froehlich} Fr\"ohlich model}

Here we consider an application of the above theory to the Fr\"ohlich model with Hamiltonian
\begin{align}
\hat{H} = \sum_{n\k} \epsilon_{n\k} c_{n\k}^{\dag} c_{n\k} + \sum_{\q\lambda} \hbar \omega_{\q\lambda} \Big( a_{\q\lambda}^{\dag} a_{\q\lambda} + \f{1}{2} \Big) + \hat{H}_1 {,}
\end{align}
where the electron-phonon interaction is 
\begin{align}
\hat{H}_1 &= \sum_{nm\k \q\lambda} g_{m\k+\q, n\k,\lambda} c_{m\k+\q}^{\dag} c_{n\k} (a_{\q\lambda} + a_{-\q\lambda}^{\dag}) {.}
\end{align}
We further simplify this to a single free-electron-like band and a single phonon mode in one dimension, i.e.
\begin{align}
\hat{H} &= \sum_{k} \epsilon_k c_{k}^{\dag} c_{k} + \sum_{q} \hbar \omega_{q} \Big( a_{q}^{\dag} a_{q} +\f{1}{2} \Big) \nn \\
&\quad + \sum_{kq} g_{k+q,k} c_{k+q}^{\dag} c_{k} (a_{q} + a_{-q}^{\dag}) {.} \label{eq:froehlich}
\end{align}
The electronic states are denoted as $|\psi_k\rangle = c_k^{\dag} |\mathrm{elec\;vac}\rangle$ and a general multiphonon state as
\begin{align}
|n_{q_1} n_{q_2} \ldots n_{q_m}\rangle = \underbrace{a_{q_1}^{\dag} \ldots a_{q_1}^{\dag}}_{n_{q_1}} \cdots \underbrace{a_{q_m}^{\dag} \ldots a_{q_m}^{\dag}}_{n_{q_m}} |\mathrm{phon\;vac}\rangle {,}
\end{align}
where $n_q$ stands for the number of phonons in mode $q$.  

\subsection{\label{ssec:density} First-order conditional density}

We will apply perturbation theory for weak electron-phonon coupling $g_{k+q,k}$.  The ground state for $g_{k+q,k}=0$ will be denoted as 
\begin{align}
|\Psi_0\rangle &= \prod_{k}^{\rm occ} c_{k}^{\dag} |0\rangle \otimes |\chi_0\rangle \nn \\
&= |\rm{FS}\rangle \otimes |\chi_0\rangle {,}
\end{align}
where $|\rm{FS}\rangle$ is the electronic Fermi sea and $|\chi_0\rangle$ is the vibrational ground state in Eq.~(\ref{eq:chi0}).  The excited states to which $|\Psi_0\rangle$ couples under $\hat{H}_1$ will be denoted as 
\begin{align}
|\Psi_{k,-q}\rangle = c_{k+q}^{\dag} c_{k} |\mathrm{FS}\rangle \otimes a_{-q}^{\dag} |\chi_0\rangle {.} 
\label{eq:excited}
\end{align}
The first-order contribution to the wave function is
\begin{align}
|\Psi^{(1)}\rangle = \sum_{kq} \f{g_{k+q, k} f_k(1-f_{k+q})}{\epsilon_{k} - \epsilon_{k+q} - \hbar\omega_{q}} | \Psi_{k,-q}\rangle {.}
\end{align}
In the $U$-coordinate representation, the wavefunction is % to first order is
\begin{align}
&| \Psi(U)\rangle \approx |\mathrm{FS}\rangle \langle U | \chi_0\rangle \nn \\
&\quad+ \sum_{kq} \f{g_{k+q,k}f_k(1-f_{k+q})}{\epsilon_{k} - \epsilon_{k+q} - \hbar\omega_{q}} c_{k+q}^{\dag} c_{k} |\mathrm{FS}\rangle \otimes \langle U | a_{-q}^{\dag} | \chi_0\rangle {.}
\end{align}

Now we use the exact factorization method to derive the conditional electronic density to first order in $g$.  The nuclear wave function is
\begin{align}
|\chi(U)|^2 &= \langle \Psi(U) | \Psi(U)\rangle {,} 
\label{eq:chi:U}
\end{align}
where the inner product is on the electronic Hilbert space only.  From the conditional electronic wave function 
\begin{align}
|\Phi_U\rangle &= \f{| \Psi(U)\rangle}{\chi(U)} \nn \\
&\approx |\mathrm{FS}\rangle \nn \\
&+ \sum_{kq} \f{g_{k+q,k}f_k(1-f_{k+q})}{\epsilon_{k} - \epsilon_{k+q} - \hbar\omega_{q}} \f{\langle U | a_{-q}^{\dag} | \chi_0\rangle}{\langle U | \chi_0\rangle} c_{k+q}^{\dag} c_{k} |\mathrm{FS}\rangle  \nn \\
&= |\mathrm{FS}\rangle + \underbrace{\sum_{kq} \f{g_{k+q,k}f_k(1-f_{k+q})}{\epsilon_{k} - \epsilon_{k+q} - \hbar\omega_{q}} \f{U_q}{L_q} c_{k+q}^{\dag} c_{k} |\mathrm{FS}\rangle}_{|\Phi_U^{(1)}\rangle}  {,}
\end{align}
we obtain the zeroth-order and first-order contribution to the conditional electronic density 
\begin{align}
n_U^{(0)}(r) &= \sum_k f_k |\psi_k(r)|^2 \nn \\
n_U^{(1)}(r) &= 2\mathrm{Re} \sum_{kq} \f{g_{k+q,k} f_k (1-f_{k+q})}{\epsilon_{k}-\epsilon_{k+q} - \hbar\omega_{q}} \f{U_q}{L_q} \psi_{k}^*(r) \psi_{k+q}(r) {.} \label{eq:density:firstorder}
\end{align}
$n_U^{(1)}(r)$ encodes how the conditional density is perturbed by the electron-phonon interaction.

\subsection{\label{ssec:geometric} Geometric correction}

As a preliminary step, we expand $|\chi(U)|^2$ as 
\begin{align}
|\chi(U)|^2 &= \langle \Psi^{(0)} | \Psi^{(0)} \rangle + \langle \Psi^{(1)} | \Psi^{(1)}\rangle \nn \\
&\quad+ 2 \mathrm{Re} \langle \Psi^{(0)} | \Psi^{(2)}\rangle + \mathcal{O}(g^4) {.}
\end{align}
The series has only even contributions.  Choosing a gauge in which $\chi$ is real, we can write $\chi(U) = \chi^{(0)}(U) + \chi^{(2)}(U) + \mathcal{O}(g^4)$ with 
\begin{align}
\chi^{(0)}(U) &= \prod_{q} \LB \f{M\omega_q}{\pi\hbar} \RB^{\f{1}{4}} \exp\LB -\f{U_q U_{-q}}{4L_q^2}\RB \nn \\
\f{\chi^{(2)}(U)}{\chi^{(0)}(U)} &= \sum_{q} |\beta_q|^2 \bigg( \f{|U_q|^2}{L_q^2}  - 1 \bigg) {,}
\end{align}
where
\begin{align}
|\beta_q|^2 &= \sum_k \f{|g_{k+q,k}|^2f_k (1-f_{k+q})}{(\epsilon_{k} - \epsilon_{k+q} - \hbar\omega_{q})^2} {.} \label{eq:beta:definition}
\end{align}
The conditional electronic wavefunction to third order is
\begin{align}
|\Phi_U\rangle &\approx \underbrace{\f{|\Psi^{(0)}\rangle}{\chi^{(0)}(U)}}_{\Phi_U^{(0)}} + \underbrace{\f{|\Psi^{(1)}\rangle}{\chi^{(0)}(U)}}_{\Phi_U^{(1)}} + \underbrace{\f{|\Psi^{(2)}\rangle}{\chi^{(0)}(U)} - \f{|\Psi^{(0)}\rangle}{\chi^{(0)}(U)} \f{\chi^{(2)}(U)}{\chi^{(0)}(U)} }_{\Phi_U^{(2)}} \nn \\[0.2cm]
&\quad+ \underbrace{\f{|\Psi^{(3)}\rangle}{\chi^{(0)}(U)} - \f{|\Psi^{(1)}\rangle}{\chi^{(0)}(U)} \f{\chi^{(2)}(U)}{\chi^{(0)}(U)} - \f{|\Psi^{(0)}\rangle}{\chi^{(0)}(U)} \f{\chi^{(3)}(U)}{\chi^{(0)}(U)}}_{\Phi_U^{(3)}} {.} \label{eq:expansion} 
\end{align}
From Eq.~(\ref{eq:expansion}), we find that the only contribution to $\mathcal{E}_{geo}$ through second order comes from the term
\begin{align}
\sum_{q} \bigg< \f{\partial \Phi_U^{(1)}}{\partial U_{q}} \bigg| \f{\partial \Phi_U^{(1)}}{\partial U_{q}} \bigg> 
&= \sum_{q} \f{|\beta_q|^2}{L_q^2} {.} \label{eq:factor}
\end{align}

\subsection{\label{ssec:KS} Generalized Kohn-Sham system}

The generalized KS potential 
\begin{align}
\hat{v}_s = \hat{v}_{en} + \hat{v}_{hxc} + \hat{v}_{geo} 
\end{align}
depends parametrically on $U_{q}$, e.g. in the potential
\begin{align}
v_{en}(r,U) = -\sum_{\mu} \f{Z_{\mu} e^2}{4\pi\epsilon_0|r-R_{\mu}|} {,}
\end{align}
the atomic coordinates $\{R_{\mu}\}$ are implicit functions of the phonon amplitudes $U=\{U_{q}\}$.   

We now apply perturbation theory to the KS system to see how it reproduces the results of the previous sections, particularly Eq.~(\ref{eq:density:firstorder}).  The KS potential is expanded as
\begin{align}
\hat{v}_s = \hat{v}_{en}^{(0)} + \hat{v}_{en}^{(1)} + \hat{v}_{geo}^{(1)} + \hat{v}_{geo}^{(2)} + \mathcal{O}(g^3) {,}
\end{align}
where the superscript denotes the order in powers of $U_q$ (or, equivalently, in powers of $g$), $\hat{v}_{en}^{(0)}$ is the potential at the equilibrium atomic coordinates $R_{\mu}^{(0)}$, and $\hat{v}_{hxc}^{(n)}=0$ for all $n$.  The unperturbed potential leads to the zeroth-order KS orbitals $\psi_k(r)$ through solution of the unperturbed KS equation 
\begin{align}
\left[ -\f{\hbar^2 \nabla^2}{2m_e} + v_{en}^{(0)}(r) \right] \psi_k(r) = \epsilon_k \psi_k(r) {.}
\end{align}
The perturbations $\hat{v}_{en}^{(1)}$, $\hat{v}_{geo}^{(1)}$ and $\hat{v}_{geo}^{(2)}$ are defined by their matrix elements [cf.~Eqs.~(\ref{eq:vgeo:offdiag}) and (\ref{eq:vgeo:diag})]
\begin{align}
\langle \psi_{k+q} | \hat{v}_{en}^{(1)} | \psi_k\rangle &= g_{k+q,k} \f{U_q}{L_q} \\
\langle \psi_{k+q} | \hat{v}_{geo}^{(1)} | \psi_k\rangle &= \hbar \omega_q U_q \bigg< \psi_{k+q} \bigg| \f{\partial\psi_k}{\partial U_{q}} \bigg> \label{eq:vgeo1:froehlich} \\
\langle \psi_{k} | \hat{v}_{geo}^{(2)} | \psi_k\rangle &= \f{\hbar^2}{2M} \bigg< \psi_k \bigg| \f{\partial^2 \psi_k}{\partial U_{-q} \partial U_q} \bigg> {.} \label{eq:vgeo2:froehlich}
\end{align}
To proceed, we need to recognize that the right-hand side of Eq.~(\ref{eq:vgeo1:froehlich}) is itself dependent on $\hat{v}_{geo}^{(1)}$.  From perturbation theory applied to the KS equation, we obtain 
\begin{align}
\bigg< \psi_{k+q} \bigg| \f{\partial\psi_k}{\partial U_{q}} \bigg> &\approx \bigg< \psi_{k+q} \bigg| \f{\partial\psi_k^{(1)}}{\partial U_{q}} \bigg> \nn \\
&= \f{\partial}{\partial U_q} \f{\langle \psi_{k+q} | \hat{v}_{en}^{(1)} + \hat{v}_{geo}^{(1)} | \psi_k \rangle}{\epsilon_k-\epsilon_{k+q}} \nn \\
&= \f{g_{k+q,k}}{\epsilon_k-\epsilon_{k+q}} \f{1}{L_q} + \f{\partial_{U_q}\langle \psi_{k+q} | \hat{v}_{geo}^{(1)} | \psi_k \rangle}{\epsilon_k-\epsilon_{k+q}} \label{eq:vgeo:rhs} {.}
\end{align}
Equations~(\ref{eq:vgeo1:froehlich}) and (\ref{eq:vgeo:rhs}) lead to the following differential equation for $G\equiv\langle \psi_{k+q} | \hat{v}_{geo}^{(1)} | \psi_k\rangle$:
\begin{align}
\f{\partial G}{\partial U_q} = \f{\epsilon_k - \epsilon_{k+q}}{\hbar\omega_q} \f{G}{U_q} - \f{g_{k+q,k}}{L_q} {.}
\end{align}
We choose the particular solution  
\begin{align}
G = \langle \psi_{k+q} | \hat{v}_{geo}^{(1)} | \psi_k\rangle = g_{k+q,k} \f{U_q}{L_q} \f{\hbar\omega_q}{\epsilon_k - \epsilon_{k+q} - \hbar\omega_q} {.}
\end{align}
Using the first-order KS orbitals
\begin{align}
\psi_k^{(1)}(r) &= \sum_{q} \f{\langle \psi_{k+q} | \hat{v}_{en}^{(1)} + \hat{v}_{geo}^{(1)} | \psi_k \rangle}{\epsilon_k-\epsilon_{k+q}} \psi_{k+q}(r) \nn \\
&= \sum_{q} \f{g_{k+q,k}}{\epsilon_k-\epsilon_{k+q}} \f{U_q}{L_q} \bigg[ 1 + \f{\hbar\omega_q}{\epsilon_k-\epsilon_{k+q}-\hbar\omega_q} \bigg] \psi_{k+q}(r) \nn \\
&= \sum_{q} \f{g_{k+q,k}}{\epsilon_k-\epsilon_{k+q}-\hbar\omega_q} \f{U_q}{L_q} \psi_{k+q}(r) {,}
\end{align}
we immediately recover the result in Eq.~(\ref{eq:density:firstorder}), namely 
\begin{align}
n_U^{(1)}(r) &= 2\mathrm{Re} \sum_k^{occ} \psi_k^{*}(r) \psi_k^{(1)}(r) \nn \\
&=2\mathrm{Re} \sum_{kq} \f{g_{k+q,k} f_k (1-f_{k+q})}{\epsilon_{k}-\epsilon_{k+q} - \hbar\omega_{q}} \f{U_q}{L_q} \psi_{k}^*(r) \psi_{k+q}(r){.} \label{eq:density:firstorder:2}
\end{align}
Thus, the generalized KS system with our nonadiabatic functional approximation reproduces the exact linear response density.  {\it Remarkably, the nonadiabatic potential $\hat{v}_{geo}$ has the effect of inserting $\hbar\omega_q$ into the denominator, thus recovering the expected nonadiabatic correction.}

To determine $\langle \psi_{k} | \hat{v}_{geo}^{(2)} | \psi_k\rangle$ in Eq.~(\ref{eq:vgeo2:froehlich}), we first use perturbation theory to show that
\begin{align}
&\bigg< \psi_k \bigg| \f{\partial^2 \psi_k}{\partial U_{-q} \partial U_{q}} \bigg> \approx \bigg< \psi_k \bigg| \f{\partial^2 \psi_k^{(2)}}{\partial U_{-q} \partial U_{q}} \bigg> \nn \\
&\quad = -\f{1}{2}\f{\partial}{\partial U_{-q}} \f{\partial}{\partial U_q} \sum_p \f{\langle \psi_k | \hat{h}^{(1)} | \psi_{k+p}\rangle \langle \psi_{k+p} | \hat{h}^{(1)} | \psi_k\rangle}{(\epsilon_k - \epsilon_{k+p})^2} \nn \\
&\quad = -\f{|g_{k+q,k}|^2}{(\epsilon_k-\epsilon_{k+q}-\hbar\omega_q)^2} \f{1}{L_q^2} {,}
\end{align}
which implies 
\begin{align}
\langle \psi_{k} | \hat{v}_{geo}^{(2)} | \psi_k\rangle = -\hbar\omega_q \f{|g_{k+q,k}|^2}{(\epsilon_k-\epsilon_{k+q}-\hbar\omega_q)^2} {.} \label{eq:vgeo2}
\end{align}
Finally, is easy to show that 
\begin{align}
\sum_{kq} f_k \bigg< \f{\partial \psi_{k}^{(1)}}{\partial U_{q}} \bigg| \f{\partial \psi_{k}^{(1)}}{\partial U_{q}} \bigg> 
\end{align}
reproduces Eq.~(\ref{eq:factor}).

\subsection{Electronic band structure renormalization}

The first-order correction to the KS eigenvalues 
\begin{align}
\epsilon^{(1)}_k &= \langle \psi_k | \hat{v}_{en}^{(1)} + \hat{v}_{geo}^{(1)} | \psi_k \rangle 
\end{align}
vanishes since $\hat{v}_{en}^{(1)}$ and $\hat{v}_{geo}^{(1)}$ are off-diagonal.  

The electronic velocity renormalization (the ``wiggle'') at the Fermi energy appears in the second-order correction 
\begin{align}
\epsilon^{(2)}_k &= \sum_{q} \bigg[ \f{|\langle \psi_{k+q} | \hat{v}_{en}^{(1)} + \hat{v}_{geo}^{(1)} | \psi_k \rangle|^2}{\epsilon_k - \epsilon_{k+q}}  + \langle \psi_k | \hat{v}_{geo}^{(2)} | \psi_k\rangle \bigg] \nn \\
&= \sum_{q} \f{|g_{k+q,k}|^2}{\epsilon_k - \epsilon_{k+q} - \hbar\omega_q} \f{|U_q|^2}{L_q^2} {.}
\end{align}
To obtain the observable perturbation we average over $U$ using $|\chi(U)|^2$ as a weighting function.  The final result
\begin{align}
\overline{\epsilon^{(2)}_k}  &= \int |\chi(U)|^2 \epsilon^{(2)}_k(U) dU \nn \\
&= \sum_{q} \f{|g_{k+q,k}|^2}{\epsilon_k - \epsilon_{k+q} - \hbar\omega_q} {.} \label{eq:epsilon2}   
\end{align}
agrees with real part of the Fan-Migdal self-energy\cite{fan1951} $\mathrm{Re}\Sigma_{nnk}^{FM}(\epsilon_k)$ at $T=0$ and therefore encodes the correct electronic velocity renormalization.  

A second-order electron-phonon interaction, called the Debye-Waller term, provides another contribution to electronic band structure renormalization.\cite{antoncik1955,allen1976,allen1981}  We do not consider it here, as it is not present in the Fr\"ohlich model, although in real materials its contribution can be of the same order as the Fan-Migdal contribution.

\section{\label{sec:conclusions} Conclusions}

Exact factorization-based DFT has been applied to interacting electrons and phonons in solids.  The equations to be solved are (i) a generalized KS equation with a nonadiabatic Hartree-exchange-correlation potential that depends on the nuclear wavefunction $\chi$ and (ii) a nuclear Schr\"odinger equation with a beyond-BO potential energy surface and induced vector potential.  Exact phonons are defined from the harmonic expansion of the nuclear Schr\"odinger equation without additional approximations.

We have proposed an approximation strategy in which nonadiabatic contributions to the KS potential and nuclear PES appear as simple additive corrections.  For the Fr\"ohlich model, the self-consistent solution of (i) and (ii) within our approximation recovers the exact electron-phonon-induced first-order density response and second-order electronic band structure renormalization.  This suggests that we can obtain good results for electron-phonon effects in real materials by adding these nonadiabatic corrections to existing DFT functionals such as the LDA and GGA.

Subjects for future work are the formulation of a finite temperature theory and an investigation of the simultaneous effects of electron-electron and electron-phonon interactions, which, in principle, can be described exactly through the nonadiabatic Hartree-exchange-correlation potential $v_{hxc}^{EF}$.  Lastly, the formalism introduced here provides an efficient methodology for predicting the effect of lattice degrees of freedom on geometric and topological properties of electronic Bloch states, such as the macroscopic polarization and topological invariants.

\begin{acknowledgments}
   R.~R. thanks P.~B.~Allen for comments on the manuscript and R. {van Leeuwen} for discussions. C.~R.~P.~thanks Consejo Nacional de Investigaciones Cient\'ificas y T\'ecnicas (CONICET) for partial financial support, grant PIP 2014-2016, and ANCyT under grant PICT 2016-1087.
\end{acknowledgments}

\bibliography{bibliography}

\end{document}